\documentstyle[psfig,11pt,a4]{article}

\newcommand{\be}{\begin{equation}}
\newcommand{\ee}{\end{equation}}
\newcommand{\bea}{\begin{eqnarray}}
\newcommand{\eea}{\end{eqnarray}}

\begin{document}

\begin{titlepage}

\begin{flushright}
{\tt
    gr-qc/0505110}
 \end{flushright}

\bigskip%\vskip3mm

\begin{center}

 {\LARGE {\bf Einstein-Planck Formula, Equivalence Principle and
Black Hole Radiance}}

\bigskip
\bigskip\bigskip
 {\bf Alessandro  Fabbri}$^a$ \footnote{e-mail:
 fabbria@bo.infn.it} and
{\bf  Jos\'e Navarro-Salas}$^b$ \footnote{email:
jnavarro@ific.uv.es}

\end{center}

\bigskip%

\footnotesize \noindent
 {\it a) Dipartimento di Fisica dell'Universit\`a di Bologna and INFN
 sezione di Bologna, Via Irnerio 46,  40126 Bologna, Italy}
 \newline
{\it  b) Departamento de F\'{\i}sica Te\'orica and
    IFIC(CSIC), Universidad de Valencia, Dr. Moliner 50,
        Burjassot-46100, Valencia, Spain}

\bigskip\vskip10mm

\bigskip

\bigskip\vskip2mm

\begin{center}
{\bf Abstract}
\end{center}

The presence of gravity implies corrections to the Einstein-Planck
formula $E=h \nu$. This gives hope that the divergent blueshift in
frequency, associated to the presence of a black hole horizon,
could be smoothed out for the energy. Using simple arguments based
on Einstein's equivalence principle we show that this is only
possible if a black hole emits, in first approximation, not just a
single particle, but thermal radiation.

\bigskip

\end{titlepage}

\newpage

One hundred years ago Einstein proposed \cite{einstein05a} an
explanation for the photoelectric effect  putting forward the
revolutionary relation $E= h \nu$ suggested by Planck
\cite{planck} five years before. The quantum nature of radiation
not only emerged in its emission by matter, as suggested by the
spectrum of blackbody radiation; light is also transmitted and
absorbed by matter in quanta. A particularly important virtue of
the Einstein-Planck formula is that it is compatible with the
theory of Special Relativity \cite{einstein05b}, also proposed by
Einstein in the same miraculous  year $1905$. The fundamental
relation $E= h \nu$ turned out to be compatible, as stressed in
\cite{de broglie}, with the new emerging view of spacetime, which,
in contrast, was closely tied to the Maxwell wave-view of light. A
Lorentz transformation along the $x$ axis with velocity $v$
modifies the phase $\nu (t+x/c)$ of a plane wave, travelling in
the opposite direction, to $\nu' (t'+x'/c)$, where ${\nu}' =
\nu\sqrt\frac{1+v/c}{1-v/c}$. This Doppler shift exactly coincides
with that relating the energy $E$ of a massless particle in the
inertial frame $(t,x)$ (i.e., the time component of the
four-vector $cp^{\mu}$) with the corresponding one in the frame
$(t',x')$: $E' = E\sqrt\frac{1+v/c}{1-v/c}$. Note that in order to
reach large frequencies like
 $\nu' \sim cl_{P}^{-1}$ (where
$l_{P}$ is the Planck length), with respect to an inertial
observer at rest, we would need planckian energies and therefore
extremely high velocities ($v \simeq c$).
% We are, in any case, far away
% from this boost scale in conventional physics.

However, we can {\it a priori} encounter very large frequencies
from the gravitational Doppler effect. Assuming a spherically
symmetric compact stellar object producing, in its exterior, a
Schwarzschild geometry (from now on we use geometrized units
$G=1=c$)\be
ds^2=-\left(1-\frac{2M}{r}\right)dt^2+\frac{dr^2}{\left(1-\frac{2M}{r}\right)}+r^2d\Omega
\ , \ee the relation between the frequencies measured by static observers
at $r=const.$  ($\nu'$) and at infinity ($\nu$) is
$\nu'=\frac{\nu}{\sqrt{1-\frac{2M}{r}}}$.
%where $\nu'$ is the
%frequency at the point $r$ and $\nu$ is the frequency at infinity
%$r=+\infty$.
If the gravitational source is compact enough (i.e.,
a black hole) to allow for $r \to 2M$, a pulse of radiation with
frequency $\nu$ at infinity has a history such that its frequency
measured by a free-falling observer that falls from rest at $(t_0,
r_0)$ can grow up without bound at that point \be \label{nu'nu}
\nu'=\frac{\nu}{\sqrt{1-\frac{2M}{r_0}}} \ , \ee as $r_0 \to 2M$.
If the free-falling observer is not at rest and has some inwards
radial velocity the above result is corrected by an additional
kinematical Doppler effect and the final blueshift is
 greater.  This happens when the observer  falls from rest
from infinity, in which case \be \label{nu'nubis} \nu' \approx
\frac{\nu}{ (1-\frac{2M}{r_0})}\ , \ee as the observer approaches
the horizon.

 This leads to the
paradoxical result that the emitted quanta of radiation could have
had an arbitrary large amount of energy in the vicinity of the
black hole horizon for free-falling observers. This conclusion,
however, is based on the implicit assumption that the relation
between energy and frequency is maintained, without corrections,
in the presence of gravity. In this essay we shall show, at a
heuristic level and using simple arguments based on the
equivalence principle, that finiteness of the energy outflux is
not compatible with the black hole emitting just a single
particle, but thermal Hawking radiation \cite{hawking}.

When a gravitational field is present the changes of coordinates
involved in the analysis  are not longer Lorentz transformations.
Restricting ourselves to the relevant $(t-r)$ sector we note that
the frequency simply transforms as \be \label{nu'nu2} \nu'=
\frac{du}{d\xi^-} \ \nu \ , \ee where $\xi^-$ is the locally
inertial (outgoing) null coordinate and $u=t-r^*$ ($r^*\equiv
r+2M\ln[(r-2M)/2M]$) is the corresponding one at infinity. The
quotient of frequencies $\nu'/\nu$ measures the different inertial
time rates at the emission and detection points, and this is,
indeed, how one gets  (\ref{nu'nu}) and (\ref{nu'nubis}).

%A simple calculation leads to \bea \label{2orderinertial}\xi^-&=&
%\frac{1}{eb}[U + \frac{V_{0}}{16M^2e}U^2+ O(U^3)]\\ \ \xi^+&=&b[V+
%O(V^3)] \nonumber , \eea where $U\equiv -4Me^{-u/4M}$ and $V\equiv
%4Me^{v/4M}$ are the null Kruskal coordinates and $V_{0}$
%parameterizes the point at the event horizon crossed by the
%free-falling observer. The constant $b$ parameterize the freedom
%of making arbitrary Lorentz transformations in the radial
%direction. It measures the velocity of the free-falling observer
%at the horizon with respect to  a closed static observer. The null
%coordinates  $u, v$ are locally inertial  at infinity. The first
%term in the expansion (\ref{2orderinertial}) leads immediately to
%(\ref{nu'nu}). As usual, the quotient of frequencies measures the
%different inertial time rates at the detection and emission
%points, respectively.

The energy of the quanta is constructed in a different way. To
simplify things  let us assume spherical symmetry. The energy of
a (spherical) outgoing narrow pulse of radiation can be defined
as $ E= \int \langle \Psi|T_{uu}|\Psi\rangle du$ at infinity,
whereas $E'= \int \langle \Psi|T_{\xi^-\xi^-}|\Psi\rangle d\xi^-$
at the horizon. $|\Psi\rangle$ represents the quantum state of
the radiation. Therefore the crucial point is to unravel how the
quantities
 $\langle\Psi|
T_{uu}|\Psi\rangle$ and $\langle\Psi| T_{\xi^-\xi^-}|\Psi\rangle$,
which represent the luminosity of the radiated particle (i.e., energy per
unit proper time) at different spacetime points, are related.
Note that these quantities are related to the expectation values
of the four-dimensional stress-energy tensor ${\bar {T}_{ab}}$ by
the expression ${\bar {T}_{ab}}=T_{ab}/4\pi r^2$.

We can estimate the relation between  $\langle\Psi|
T_{uu}|\Psi\rangle$ and $\langle\Psi| T_{\xi^-\xi^-}|\Psi\rangle$
neglecting the backscattering of the radiation between the
emission and detection points. Moreover, putting aside the angular
coordinates, the stress-energy tensor can be identified directly
with the corresponding normal-ordered expression in the locally
inertial frame\footnote{We neglect this way subleading
contributions coming from the spatial curvature of the metric.}
\cite{icp}\be \langle \Psi |T_{\xi^-\xi^-}|\Psi\rangle
\approx\langle \Psi |:T_{\xi^-\xi^-}:|\Psi\rangle \ . \ee
Therefore, we can apply the rules of planar conformal invariance
\cite{bpz}, and then a simple answer emerges  \be \langle
\Psi|:T_{uu}:|\Psi \rangle= \left (\frac{d\xi^-}{du}\right )^2
\langle \Psi|:T_{\xi^-\xi^-}:|\Psi\rangle -\frac{\hbar}{24\pi
}\{\xi^-, u\} \ , \ee where $ \{\xi^-,u\}=
\frac{d^3\xi^-}{du^3}/\frac{d\xi^-}{du}-\frac{3}{2}\left
(\frac{d^2\xi^-}{du^2}/\frac{d\xi^-}{du}\right )^2$ is the
Schwarzian derivative. Note that, generically, the relation should
be of the form \be \label{generaltrans}\langle \Psi
|T_{uu}|\Psi\rangle = \left (\frac{d\xi^-}{du}\right )^2\langle
\Psi |T_{\xi^-\xi^-}|\Psi\rangle + \hbar
C(u,\xi^-;g_{\mu\nu},\Psi) \, \ee where $C$, which depends on the
initial and final points, the background metric and the quantum
state, represents a correction to the flat-space formula. So we
have \be \label{E'E}E'\equiv \int\langle \Psi
|T_{\xi^-\xi^-}|\Psi\rangle d\xi^-=\int\frac{du}{d\xi^-}\langle
\Psi |T_{uu}|\Psi\rangle du + \hbar\int
\left(\frac{du}{d\xi^-}\right)^2C d\xi^-
 \ , \ee
For instance, for a one-particle (wave-packet) state of
frequency around $\nu$ and peaked about the time $u \approx u_0$
we have $E\approx h\nu$ and therefore \be
E'=\frac{du}{d\xi^-}|_{u_0}h\nu+ \dots
 \ , \ee
where the leading term fits expression (\ref{nu'nubis}). The
corrections can be worked out entirely in the approximation we are
considering. Neglecting the backscattering of the radiation we
get, for the
 one-particle state considered before,

\be \label{E'E2}E'=  \frac{du}{d\xi^-}|_{u_0}h\nu +
\frac{\hbar}{24\pi }\int \left(\frac{du}{d\xi^-}\right)^2
\{\xi^-,u\}d\xi^- \ . \ee
%\be \label{E'E2}E'=  h\nu \frac{e^{u_0/4M}}{eb} +
%\frac{\hbar}{24\pi }\int \frac{du}{d\xi^-} \{\xi^-,u\}du \ . \ee
Note that the term involving the Schwarzian derivative in
(\ref{E'E2}) turns out to be independent of the particular quantum
state.  This is due to the fact that we are considering the
``emission" point very close to the horizon. Were it not located
close to the horizon, the correction would be, in general,
state-dependent. Therefore, and just at the vicinity of the
horizon, this term can be interpreted as a vacuum energy
contribution.

We also remark that the correction codifies the fact that the
locally inertial coordinates at the two points are different, and
not related by Lorentz transformations (for them the Schwarzian
derivative vanishes and we recover the usual relation $E'=h
\nu'$). However, we must note that to evaluate  (\ref{E'E2}) we
need to know, in addition to the first derivative of the
relation $\xi^-=\xi^-(u)$ (as required to relate the frequencies), the
second and also the third derivatives. To get rid of the acceleration
of a point-particle at a given point, as invoked by Einstein's
equivalence principle, it is enough to know the first and second
derivatives. A simple calculation leads to \bea
\label{secondorder}
\xi^-&=&\frac{d\xi^-}{du}|_{u_0}[(u-u_0)-\frac{M}{2r_0^2}(u-u_0)^2+
O((u-u_0)^3)] \ , \\
\xi^+&=&\frac{d\xi^+}{dv}|_{v_0}[(v-v_0)+\frac{M}{2r_0^2}(v-v_0)^2+
O((v-v_0)^3)] \ , \eea where $v=t+r^*$ and $r_0$ is given by the
relation $(v_0-u_0)/2= r_0 + 2M\ln[(r_0-2M)/2M]$. Since now we
are working with extended objects (quantum wave packets) it should
not be surprising that higher-order conditions emerge. We have to
make use of the local Lorentz frame of a free-falling observer
\cite{mtw}. This in practice requires one  to select the normal
Riemann coordinates. Every time-like or null geodesic passing
through the preferred point $(u_0, v_0)$ is a straight line in
that frame. So the expression (\ref{secondorder}) should be
improved by adding the corresponding third order. The calculation
leads to \bea \label{thirddorder}
\xi^-&=&\frac{d\xi^-}{du}|_{u_0}[(u-u_0)-\frac{M}{2r_0^2}(u-u_0)^2
-\frac{M(r_0-3M)}{6r_0^4}(u-u_0)^3  \nonumber \\
 &+& O((u-u_0)^4)] \ . \eea
%\be \label{3orderinertial}\xi^-= \frac{1}{eb}[U +
%\frac{V_{0}}{16M^2e}U^2+ ...U^3+O(U^4)]\ . \ee
Now we can estimate the flux of energy in the vicinity of the
horizon. For points close to the horizon $r_0 \to 2M$ ($u_0 \to
+\infty$), we find that \bea \label{basicequation}\langle \Psi
|:T_{\xi^-\xi^-}:|\Psi\rangle|_{\xi^- \approx 0}
&=&\left(\frac{du}{d\xi^-}\right)^2[\langle \Psi
|:T_{uu}:|\Psi\rangle + \frac{\hbar}{24\pi } \{\xi^-,u\}]|_{u_0}
 \nonumber \\
&\approx&\left(\frac{du}{d\xi^-}\right)^2[\langle \Psi
|:T_{uu}:|\Psi\rangle|_{u_0} - L+ A(1-\frac{2M}{r_0})^2]
 \ ,  \ \ \ \ \ \ \eea
where $L=\frac{\hbar}{768\pi M^2}$ and $A=\frac{\hbar}{128 \pi
M^2}$. Note that, remarkably, the value obtained for $L$ is
sensitive to all the three terms explicitly written in the expansion
(\ref{thirddorder})\footnote{When backscattering is included the constant $L$ and
also $A$ are modified by grey-body factors.}.

Since the energy flux of a single particle at infinity $\langle
\Psi |:T_{uu}:|\Psi\rangle|_{u_0}$ is  peaked around the point
$u_0$  it can never balance the constant term $L$ in
(\ref{basicequation}). This is in fact a realization of
Heisenberg's uncertainty principle. Therefore,
although the relation between energy and frequency is modified (see eq. (\ref{E'E2}))
the final result is qualitatively unchanged because
the value of
$\langle \Psi |:T_{\xi^-\xi^-}:|\Psi\rangle$ at the horizon is
necessarily divergent. We can interpret this outcome by saying that
if we require finiteness of the energy flux then this implies
that a black hole can never radiate out a single particle. This
conclusion still holds for a generic many particle state in the
Fock space. Only when the emitted radiation gives a (constant)
{\it thermal} luminosity $\langle \Psi|:T_{uu}:|\Psi\rangle =
\frac{\pi}{12\hbar} T^2$, with Hawking temperature $T=\hbar/8\pi
M$, both constant terms can be summed up to produce a finite
energy flux at the black hole horizon. Similar arguments apply for
the finiteness of the correlation $\langle\Psi|:T_{\xi^-\xi^-}(1):
:T_{\xi^-\xi^-}(2):|\Psi\rangle$ at the horizon, which also
requires the thermal correlation for $\langle\Psi|:T_{uu}(1):
:T_{uu}(2):|\Psi\rangle$.

We finally stress that if one neglects the third order term in
(\ref{thirddorder}) and keeps only the usual first and second
terms as requested for the definition of locally inertial
coordinates, the divergent blueshift
cannot be removed. The thermal nature of the radiation is closely
related to the explicit form of the ``normal coordinates" as one
approaches the classical horizon. Finally, let us mention that
backreaction effects will certainly
modify the location and structure of the horizon and therefore the
associated radiation. It may be interesting to explore
their physical implications within the picture offered in this essay.

%COMENTARIO. Con una metrica del tipo \be
%ds^2=-(1-\frac{2M(u)}{r})du^2-2dudr\ee Si se impone conservacion
%de energia\be \frac{dM(u)}{du}=-<T_{uu}> \ee

%y que sea finito $<T_{\xi^-\xi^-}>$ en $r=2M(u)$ se obtendria otra
%formula para la radiacion. Tal vez diera una expression
%interesante ???. En Parik-Wilczek consiguen dar una correccion a
%la radiacion termica utilizando conservacion de energia.

\end{document}